\documentclass[prd,showpacs,amsmath,amssymb,nofootinbib,superscriptaddress]{revtex4}
\usepackage{array,mathtools,amssymb,booktabs}
\usepackage[english]{layout}
\usepackage[english]{babel}
\usepackage{bm}
\usepackage{graphicx}
\usepackage{hhline}
\newcommand{\Od}{{\cal O}}

\newcommand{\tintq}{T\sum_n \int\frac{d^3 \vec{q}}{(2\pi)^3}}

\newcommand{\pif}{\bm{\pi}}
\newcommand{\cond}{\langle \bar q q \rangle}

\newcommand{\quarkcorlT}{\langle {\cal T} (\bar q q)_l(x) (\bar q
q)_l(0)\rangle_{T}}
\newcommand{\intT}{\int_0^\beta d\tau \int d^3 \vec{x}}

\begin{document}

\title{Chiral Symmetry Restoration for the large-$N$ pion gas}
\author{Santiago Cort\'es}
\email{js.cortes125@uniandes.edu.co}
\affiliation{Departamento de F\'{\i}sica, Univ. de Los  Andes, 111711 Bogot\'a, Colombia.}
\author{A.~G\'omez Nicola}
\email{gomez@ucm.es}
\affiliation{Departamento de F\'{\i}sica
Te\'orica II. Univ. Complutense. 28040 Madrid. Spain.}
\author{John Morales}
\email{jmoralesa@unal.edu.co}
\affiliation{Departamento de F\'{\i}sica, Univ. Nacional de Colombia, 111321 Bogot\'a, Colombia.}
\date{\today}

\begin{abstract}
We analyze chiral restoration within the $O(N+1)/O(N)$ Non-Linear Sigma Model for large $N$ as an effective theory for low-energy QCD at finite temperature $T$. The free energy is constructed diagramatically to $\Od(TM^3)$ in the pion mass expansion, which allows to derive the quark condensate and the scalar susceptibility in the chiral limit. At this order,  we do not have to deal with renormalization, neither from divergences from mass tadpoles nor from those of higher order loop contributions. Our results for the critical behaviour are consistent with expectations from lattice analysis and with previous works where the susceptibility is saturated by the thermal $f_0(500)$ pole.

\vspace*{0.5cm}
\end{abstract}

\maketitle

\section{Introduction} 

The study of hadronic properties at finite temperature $T$ is one of the theoretical ingredients needed to understand the behaviour of  matter created in Relativistic Heavy Ion Collision experiments, such as those  in RHIC and LHC (ALICE). In particular, the QCD transition involving chiral symmetry restoration and deconfinement plays a crucial role, as it is clear from the many recent advances of lattice groups in the study of the phase diagram and other thermodynamical properties \cite{Aoki:2009sc,Ejiri:2009ac,Bazavov:2011nk,Bhattacharya:2014ara}.  For vanishing baryon chemical potential, the QCD transition is a crossover for  2+1 flavours with physical quark masses, the transition temperature being about  $T_c\sim$ 150 - 160 MeV, determined by the vanishing behaviour of the quark condensate and the peak of the scalar susceptibility. In the chiral limit it becomes   a second-order phase transition consistent with the $O(4)$-model universality class  \cite{Pisarski:1983ms,Berges:2000ew}, which is supported in lattice simulations by  the mass and temperature scaling of thermodynamical quantities as well as chiral partner degeneration    \cite{Bazavov:2011nk,Ejiri:2009ac}. The expected reduction in the transition temperature from the physical mass case to the chiral limit one based on those analysis is about 15-20$\%$ \cite{Bazavov:2011nk}. 

From the theoretical side, it  is therefore important to provide solid analysis of this chiral restoration pattern based on effective theories, given the limitations of perturbative QCD at those temperature scales.  Such effective description should start from a proper understanding of the lightest component, i.e. the pion gas. Pions are actually the most abundant particles after a Heavy Ion Collision and most of their properties from hadronization to thermal freeze-out can be reasonably described within the temperature range where effective theories are applicable. In fact, approaches based on effective theories for the lightest mesons provide a good description of the physics involved, especially in what concerns the effect of the lightest resonances, as we discuss below. A more accurate treatment of thermodynamic quantities near $T_{c}$ would require including heavier degrees of freedom, which can be efficiently achieved through the  Hadron Resonance Gas framework \cite{Tawfik:2005qh,Huovinen:2009yb}.

A systematic and model-independent framework that takes into account the relevant light meson degrees of freedom and their interactions is Chiral Perturbation Theory (ChPT) \cite{Gasser:1983yg}. The effective ChPT Lagrangian is constructed as a derivative and mass expansion  ${\mathcal L}={\mathcal L}_{p^2}+{\mathcal L}_{p^4}+\dots$, where $p$ denotes generically a meson energy scale compared to the chiral scale $\Lambda_{\chi}\sim$ 1 GeV. The lowest order Lagrangian ${\mathcal L}_{p^2}$ is the Non-linear Sigma Model (NLSM).  Thus,  chiral restoring behaviour is qualitatively obtained within ChPT through the vanishing quark condensate for different orders \cite{Gerber:1988tt}, although a critical  description is not obtained in the chiral limit.  Despite the  model-independent character of the ChPT predictions for chiral restoration,  the low-$T$ nature of this theory implies  a continuous behaviour for order parameters and susceptibilities, even in the chiral limit. Thus, the quark condensate is a
dropping continuous function around the critical point and the scalar susceptibility grows also continuously.  In addition, ChPT is unable to describe resonant states, which play a crucial role in the description of the hadronic medium. Some of these limitations are improved within the unitarized framework at finite temperature \cite{GomezNicola:2002tn,Dobado:2002xf}, which provides an accurate description of several effects of interest in a Heavy-Ion environment, such as thermal resonances and transport coefficients \cite{FernandezFraile:2009mi}. It provides also a novel understanding of the role of the $\sigma/f_0(500)$ $I=J=0$ thermal pole (a broad resonant state) in chiral symmetry restoration. Thus, the scalar susceptibility  saturated with this $\sigma$-like state within the so-called Inverse Amplitude Method (IAM) unitarization, develops a maximum near $T_{c}$  \cite{Nicola:2013vma} compatible with lattice data and chiral partners in the scalar-pseudoscalar sector are understood through degeneration of correlators and susceptibilities.  The role of the $f_0(500)$ state for chiral restoration could become more complicated if its possible tetraquark component is also considered at finite temperature  \cite{Heinz:2008cv}.

A complementary approach is the large-$N$ one, where $N$ is the number of light Nambu-Goldstone Bosons (NGB). Within this framework,  the lowest order chiral effective Lagrangian for low-energy QCD will be the $O(N+1)/O(N)$ NLSM, whose corresponding symmetry breaking pattern is $O(N+1)\rightarrow O(N)$. As we have just commented, the latter is believed to take place in chiral symmetry restoration for $N=3$, since $O(4)$ and $O(3)$ are respectively isomorphic to the isospin groups $SU_L(2)\otimes SU_R(2)$ and $SU_V(2)$. In this limit, many of the features discussed above arise naturally. For instance, the pion scattering amplitude generates the $f_0(500)$ resonance in accordance with scattering and pole data. When extended at finite temperature, thermal unitarity holds exactly and the thermal pole gives rise to  a saturated scalar susceptibility diverging in the chiral limit at the critical temperature as a second-order phase transition \cite{Cortes:2015emo}. Within a  similar context, it is worth mentioning also recent large-$N$ analysis in the vector $O(N)$ model regarding the $\sigma$ spectral properties  at finite $T$ \cite{Patkos:2002xb}. 

The large-$N$ framework for the NLSM has been analyzed  in earlier works under various approximations. At $T=0$,  functional methods were developed in \cite{Appelquist:1980ae}; the scattering amplitude, including its renormalizability, was studied in the chiral limit in \cite{Dobado:1992jg} and to leading order in mass corrections in \cite{Dobado:1994fd}. At finite temperature, apart from the previously mentioned  work \cite{Cortes:2015emo}, the free energy and the quark condensate within a saddle-point approximation for the auxiliary field were analyzed in \cite{MeyersOrtmanns:1993dw}.  Other studies in the chiral limit that also utilize functional methods \cite{Bochkarev:1995gi} provide a chiral restoring analysis and the $T$-dependence of the pion decay constant \cite{Jeon:1996gn}.  A later work \cite{Andersen:2004ae} studied the NLSM as the infinite coupling limit of the $O(N)$ vector model, their results for the NLO pressure in the chiral limit not being fully consistent with those  in \cite{Bochkarev:1995gi}. Various renormalization-group studies of the critical properties of this model were compiled in \cite{zjbook}.

In this work we will analyze  the leading large-$N$ contributions to the quark condensate and the scalar susceptibility, derived diagrammatically from the partition function or the free energy.  Since we are mostly interested in the study of the critical behaviour, we will restrict to the leading order in the expansion around the chiral limit, which has the additional advantage of yielding results that are not to be renormalized. The scalar susceptibility and the  analysis of the critical behaviour are new from this work.  In addition, another motivation for the present study and a prominent difference with respect to previous analysis is that we will work directly within the diagrammatic approach to the partition function, identifying the dominant contributions  to the free energy. Thus, that approach will not introduce additional assumptions within the auxiliary field method, such as saddle point approximations, but will require  a careful evaluation of the diagrams and effective vertices involved.  In fact, for the susceptibility analysis we will deal with a resummation of an infinite set of closed ring diagrams, which  will confirm qualitatively the analysis performed previously in terms of pion scattering and the thermal $f_0(500)$ saturation. That comparison is also another motivation for the present work. We will need to calculate the large-$N$  free energy up to order $TM^3$ in order to extract properly the leading order susceptibility near the chiral limit, which actually means going beyond previous analysis of the NLSM for large $N$. In this sense, our present study aims to set up the correct diagrammatic framework for future analysis beyond the chiral limit. The paper is organized as follows: we will present our main formalism in section \ref{sec:form}, the detailed diagrammatic analysis will be performed in section \ref{sec:diag} and the results will be presented in section \ref{sec:res}, where we compare the obtained critical behaviour with previous theoretical approaches, as well as with lattice results for critical exponents.

\section{Formalism and conventions}
\label{sec:form}

We start from the Lagrangian of the nonlinear $S^N=O(N+1)/O(N)$ model with a explicit symmetry breaking term which generates the pion mass \cite{Dobado:1994fd}

\begin{eqnarray}
\mathcal{L}_{NLSM}&=&\frac{1}{2}\left[\delta_{ab}+\frac{1}{NF^{2}}\frac{\pif_{a}\pif_{b}}{1-\pif^2/NF^{2}}\right]\partial_{\mu}\pif^{a}\partial^{\mu}\pif^{b}
+NF^2M^2\sqrt{1-\frac{\pif^2}{NF^2}} \nonumber\\
&=&NF^2M^2+\frac{1}{2}\partial_{\mu}\pif^{a}\partial^{\mu}\pif^{a}-\frac{1}{2}M^2\pif^2+\frac{1}{2NF^2}\pif^a\pif^b \partial_\mu\pif_a\partial^\mu\pif_bf\left(\frac{\pif^2}{NF^2}\right)-\frac{M^2}{8NF^2}(\pif^2)^2g\left(\frac{\pif^2}{NF^2}\right),
\label{NLSM}
\end{eqnarray}
with $\pif^2=\displaystyle\sum_{a=1}^{N}\pif_{a}\pif^{a}$ and $M^2$ and $\sqrt{N}F$ are respectively the pion mass and  the pion decay constant in the chiral limit. Here we have explicitly separated the kinetic free Lagrangian and write the interaction part, with and without derivatives, in terms of the  functions

\begin{eqnarray}
f(x)&=&\frac{1}{1-x}=\sum_{k=0}^\infty x^k\nonumber ,\\
g(x)&=&-\frac{8}{x^2}\left[\sqrt{1-x}-1+\frac{x}{2}\right]=-8\sum_{k=0}^{\infty}(-1)^k \begin{pmatrix} 1/2\cr k+2 \end{pmatrix} x^k=1+\frac{x}{2}+\frac{5}{16}x^2+\cdots,
\end{eqnarray}
where $\displaystyle\begin{pmatrix} \alpha\cr n \end{pmatrix}=\frac{1}{n!}\alpha (\alpha-1)\cdots (\alpha-n+1)$. Both $f(x)$ and $g(x)$ are normalized so that the leading order in the $1/F^2$ ChPT expression ($x\rightarrow 0$) corresponds to $f=g=1$. Similar functions are used in the analysis of pion scattering in the massive case at $T=0$   within the auxiliar field method \cite{Dobado:1994fd}.

The free energy from which all the thermodynamic variables can be extracted is given by

\begin{equation}
z(M,T)=-T\lim_{V\rightarrow\infty} \frac{1}{V}\log Z(M,T), \qquad 
Z(M,T)=\int d\pif \exp\int_T \mathcal{L}[\pif],
\label{freeenergy}
\end{equation}
where $Z(M,T)$ is the QCD partition function in the pionic sector, hence expected to be dominant at low and moderate temperatures, and $\int_T\equiv\intT$, $\beta=1/T$, $\mathcal{L}[\pif]=\mathcal{L}_{NLSM}[\pif]+\dots$, the dots indicating higher order Lagrangians in derivatives and masses, which would eventually have to be included to renormalize the theory, within the same approach followed in previous works \cite{Gasser:1983yg,Gerber:1988tt,Dobado:1992jg, Dobado:1994fd,Cortes:2015emo}. 

The light quark condensate behaves as an order parameter for chiral symmetry restoration in the chiral limit:
  
  \begin{equation}
  \cond (M,T)=\frac{\partial z (M,T)}{\partial m_q}=2B_0 \frac{\partial z (M,T)}{\partial M^2},
  \label{quarkconddef}
  \end{equation}
where $m_q$ is the light quark mass, $\bar q q=\sum_{i=1}^{N_f} \bar q_i q_i$ with $N_f$ the number of light flavours and we have used the standard relation between the NGB mass and the quark mass $M^2=2B_0 m_q$, related also to the $T=0$ quark condensate in the chiral limit $\cond (T=0)=-2NF^2B_0$.

The quark condensate correlator defines the scalar susceptibility, namely:

\begin{eqnarray}
\chi_S(M,T)&=&-\frac{\partial}{\partial m_q} \cond (M,T)=-\frac{\partial^2}{\partial m_q^2}z(M,T)=-4B_0^2\frac{\partial^2}{\partial (M^2)^2}z(M,T)
\nonumber \\
&=&
\int_T{d^4x \left[\quarkcorlT-\cond^2(T)\right]}.
\label{chil4q}\end{eqnarray}

An important comment is that throughout this work, we will be interested only in the chiral limit $M\rightarrow 0^+$, which, as explained above, should capture the essential features of the chiral phase transition, both for the quark condensate and for the scalar susceptibility. However, from their  previous definitions \eqref{quarkconddef} and \eqref{chil4q}, we see that it is necessary to keep $M^2$ finite and send it to zero only after differentiation. Thus, we will consider the large-$N$ leading contribution for finite mass and then we will keep only the relevant terms in the $M^2$ expansion near the chiral limit. This is a distinctive feature with respect to  previous large-$N$ NLSM analysis at finite $T$, which analyze $z(T)$ within the limit of massless pions, but not its $M^2$ corrections \cite{Bochkarev:1995gi,Andersen:2004ae},  particularly relevant for the case of $\chi_S$, since, from its divergent nature near the transition it is indeed expected to behave as an $\Od(M^{-1})$ quantity \cite{Smilga:1995qf,GomezNicola:2012uc}, which we will also obtain  in this approach by keeping the relevant $M^3$ terms. 

In the following analysis, the free energy density will be expressed in terms of different thermal functions. Following the same notation as in \cite{Gerber:1988tt}, we define $g_k(M,T)$ satisfying $g_{k+1}(M,T)=-\frac{\partial}{\partial M^2} g_k(M,T)$ and whose expansion in  $M/T$ reads (we keep the terms relevant for this work)

\begin{eqnarray}
g_0(M,T)&=&\frac{\pi^2}{45}T^4\left[1-15\left(\frac{M}{2\pi T}\right)^2+60 \left(\frac{M}{2\pi T}\right)^3+\Od\left(\left(\frac{M}{2\pi T}\right)^4\log\left(\frac{M}{2\pi T}\right)\right)\right]\label{g0},\\
g_1(M,T)&=&\frac{T^2}{12}\left[1-3\frac{M}{\pi T}+\Od\left(\left(\frac{M}{2\pi T}\right)^2\log\left(\frac{M}{2\pi T}\right)\right)\right]\label{g1},\\
g_2(M,T)&=&\frac{T}{8\pi M}+\Od\left(\log\left(\frac{M}{2\pi T}\right)\right).\label{g2}
\end{eqnarray}

The above $g_i$ functions arise naturally from thermal parts of loop functions. Thus, defining  $G_1(M,T)=G(x=0)$ as the tadpole function with $G$ the free pion propagator, we have:

\begin{eqnarray}
G_1(M,T)&=&\tintq \frac{1}{\omega_n^2+\vert\vec{q}\vert^2+M^2}=G_1(M,0)+g_1(M,T),\nonumber\\  
G_1(M,0)&=&M^2\left[(4\pi)^{-D/2}\Gamma\left(1-\frac{D}{2}\right)\mu^{D-4}+\frac{1}{16\pi^2}\log\frac{M^2}{\mu^2}\right],
\label{G1}
\end{eqnarray}
with $\omega_n=2\pi n T$ the Matsubara frequencies, $\mu$ the renormalization ChPT scale and $\Gamma$ the Euler gamma function. We follow the same notation as in \cite{Gasser:1983yg}.

The $g_2$ function corresponds to the thermal part of the scattering loop with zero external momenta: 
\begin{equation}
G_2(M,T)=\tintq \frac{1}{\left(\omega_n^2+\vert\vec{q}\vert^2+M^2\right)^2}=-\frac{d}{dM^2} G_1(M,T)=G_2(M,0)+g_2(M,T).
\label{G2}
\end{equation}

Similarly, we define the loop function

\begin{eqnarray}
G_k(M,T)&=&\tintq 
 \frac{1}{\left(\omega_n^2+\vert\vec{q}\vert^2+M^2\right)^k}=\frac{1}{(k-1)!}\left(-\frac{d}{dM^2}\right)^{k-1}G_1(M,T) \nonumber \\
&=&G_k(M,0)+\frac{1}{(k-1)!}g_k(M,T) \nonumber\\
 &=& \frac{TM^3}{8\pi} \frac{(2k-5)!!}{(k-1)! 2^{k-2}} \frac{1}{M^{2k}}\left[1+\Od\left(\frac{M}{T}\right)\right] \qquad \mbox{for} \ k\geq 3,
\label{Gk}
\end{eqnarray}
  where we have extracted the leading order in the $T/M$ expansion, from the asymptotic expansion \eqref{g1}.

\section{Large-$N$  and mass expansion}
\label{sec:diag}

\begin{figure}
\centering
\includegraphics[scale=0.9]{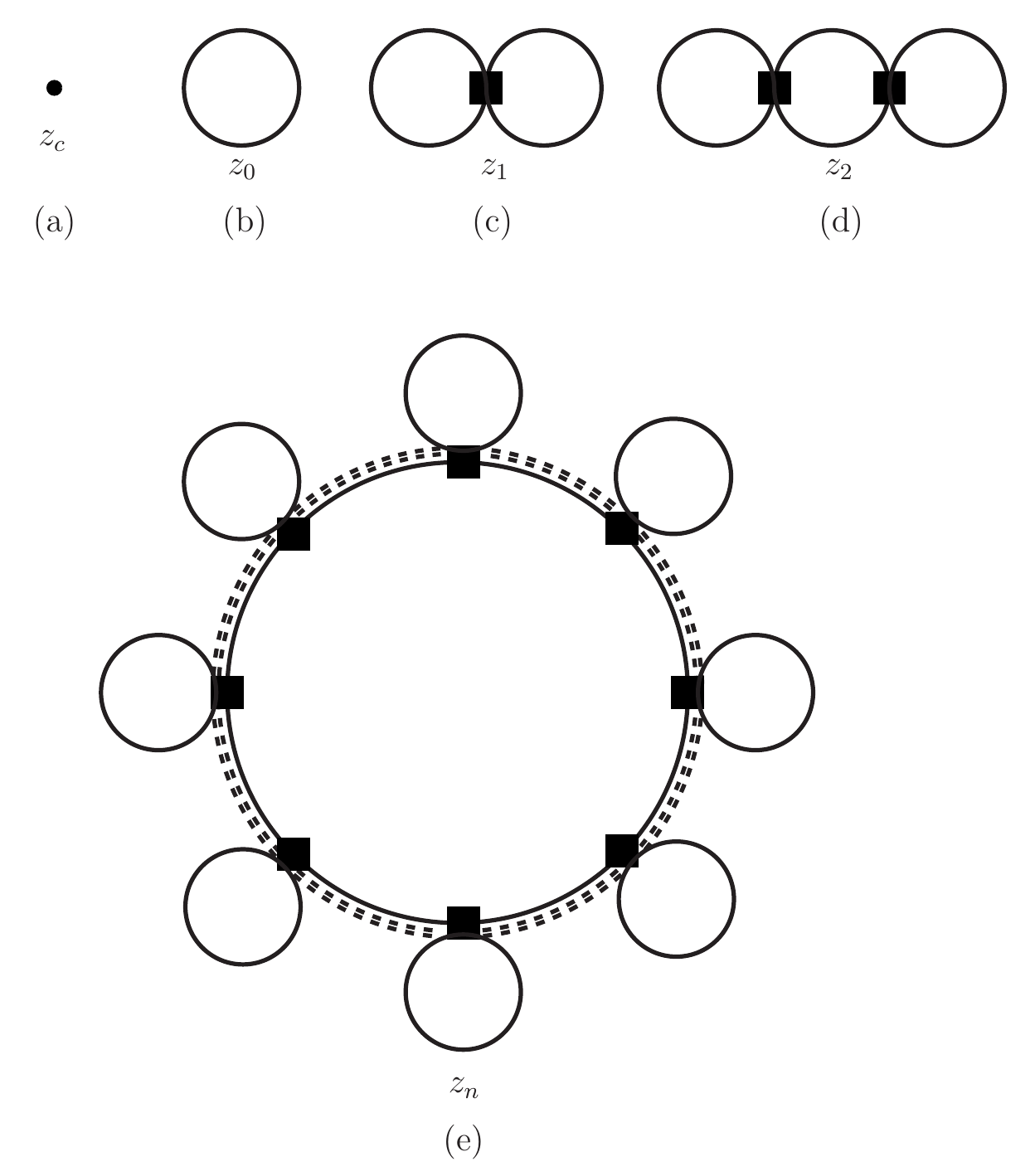}
\caption{Diagrams contributing to the free energy to leading order in $N$ and up to $\Od(TM^3)$. The black squares denote the effective vertex depicted in Figure \ref{fig:massvertex}. The different contributions (a)-(e) are explained in the main text. The dashed lines in (e) indicate multiple insertions of this vertex along the central loop.}
\label{fig:parfun}
\end{figure}

In order to identify the leading order contribution in the large-$N$ limit for fixed mass $M^2$, it is useful to examine the different possible terms for a given number of vertices, i.e., to a certain order in the expansion of the interaction part of the Lagrangian (\ref{NLSM}) in the free energy \eqref{freeenergy}. Thus, we first extract the constant term $z_c=-NF^2M^2$ which is $\Od(N)$ and we will denote by $z_n$ the leading contribution for $n$ vertices at large $N$ for fixed $M^2$. The contribution with zero vertices is that coming from the free kinetic pion part of the NLSM Lagrangian \eqref{NLSM}; this  corresponds to  the free partition function of a $N$-component massive boson gas:

\begin{equation}
z_0(M,T)=-\frac{N}{2}\left[(4\pi)^{-D/2}\Gamma\left(-\frac{D}{2}\right)M^D+g_0(M,T)\right],
\label{z0}
\end{equation}
which is also $\Od(N)$.   The contributions $z_c$ and $z_0$ are given by diagrams (a) and (b) in Figure \ref{fig:parfun}, respectively. 

We also keep the space-time dimension $D=4-\epsilon$ in the dimensional regularization scheme. The free energy is divergent and needs renormalization, as it is discussed in detail within the ChPT expansion in \cite{Gerber:1988tt}. Renormalization within the large-$N$ framework can also be carried out by including suitable higher order Lagrangians in derivatives and masses. This is reviewed for the scattering amplitude at $T=0$ in \cite{Dobado:1992jg,Dobado:1994fd} and for $T\neq 0$ in \cite{Cortes:2015emo}. At the order considered in this work, as we will see, the dominant contribution is finite, so we will not need to implement explicitly such renormalization procedure.

\begin{figure}
\centering
\includegraphics[scale=0.9]{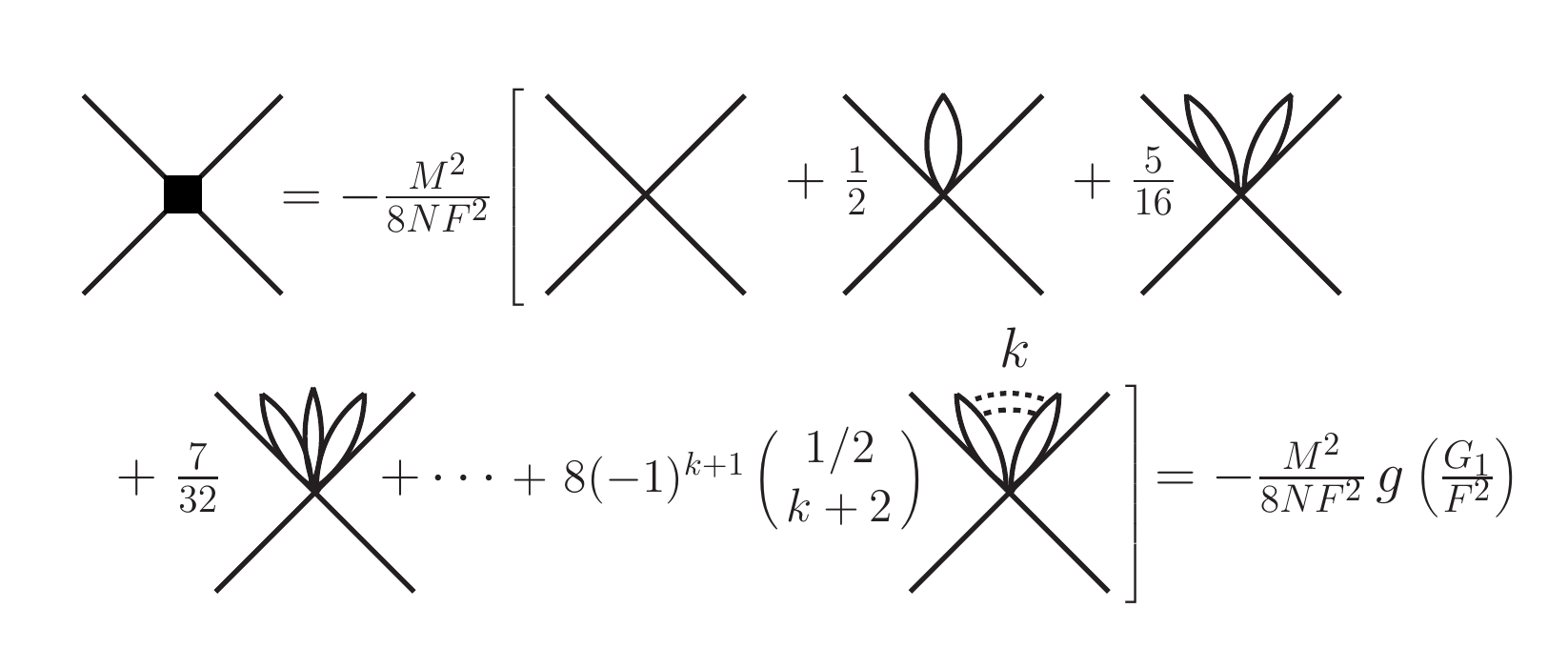}
\caption{Effective mass vertex. Dashed lines in the last diagram indicate the multiple insertions of pion tadpoles coming from contractions of pairs of extra legs.}
\label{fig:massvertex}
\end{figure}

The one-vertex contribution corresponds to the first order in the expansion of the interaction Lagrangian in (\ref{NLSM}). In the mass vertex term, i.e., the one with the $g$ function, we get the maximal contribution in $N$ by contracting all pion pairs with the same isospin indices, so that $\pif^a\pif_a\rightarrow NG_1(M,T)$ with $G_1$ in \eqref{G1}.  Therefore, we have to sum all tadpole insertions in the vertex as given by the function $g$ in the Lagrangian, which we attain by defining an effective vertex as indicated in Figure \ref{fig:massvertex}. On the other hand, one realizes that contractions involving the interaction derivative term in the Lagrangian  (the term proportional to the function $f$ in \eqref{NLSM}) give always subdominant contributions in $N$ since $\pif^a\partial_\mu\pif_a\rightarrow \partial_\mu G_1\vert_{x=0}=0$ by parity, so those contributions are $\Od(1)$ in the large-$N$ expansion of the free energy.

We  point out that the combinatoric factors shown in Figure \ref{fig:massvertex} remain the same as those obtained from the $1/N$ expansion of the mass vertex in (\ref{NLSM}). This is because our analysis, albeit diagrammatical, is not describing any specific scattering process as in previous analysis \cite{Cortes:2015emo}  but closed diagrams for the free energy, so we do not have to consider extra factors in the resummation given by the function $g(G_{1}/F^{2})$. Another significative difference with the scattering case is that here  derivative vertices do not  show up to leading order. Thus, we obtain for the large-$N$ leading one-vertex contribution to the free energy

 \begin{eqnarray}
 z_1(M,T)&=&\frac{N}{8}\frac{M^2}{F^2}G_1^2(M,T)g\left[\frac{G_1(M,T)}{F^2}\right]+\Od(N^0) \nonumber \\
&=&-NM^2F^2 \left\{ h\left(\frac{T^2}{12F^2}\right)-\frac{MT}{4\pi F^2}h' \left(\frac{T^2}{12F^2}\right)
\right\} +\Od\left[M^4\log M,N^0\right],
  \label{z1}
 \end{eqnarray}
 according to \eqref{g1} and \eqref{G1},
with $h(x)=-(1/8)x^2 g(x)$ \footnote{Note that  the function $h(x)$ corresponds to  $-g^2(x)$ in \cite{Dobado:1994fd}.}. The above contribution $z_1$ corresponds to  the connected diagram (c) in Figure \ref{fig:parfun}. The topology of this diagram is the same as in ChPT \cite{Gerber:1988tt} but here effective mass vertices enter and derivative vertices are absent.

\begin{figure}
\centering
\includegraphics[scale=0.9]{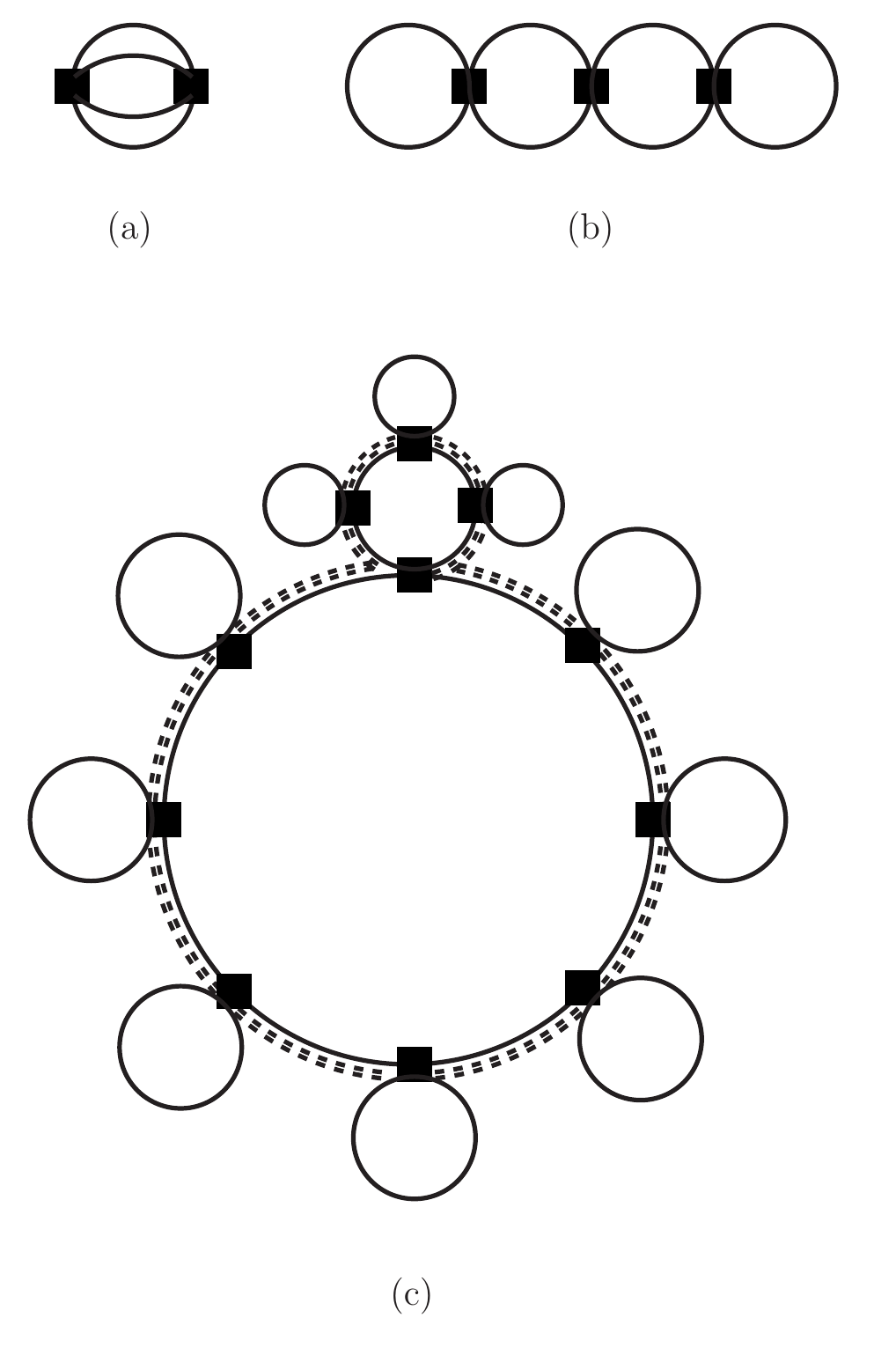}
\caption{Different examples of subleading diagrams. (a) is subleading in the large-$N$ expansion for fixed $M$, while (b) and (c) are of leading $\Od(N)$ order but subleading in the $M^2$ expansion.}
\label{fig:otherdiagrams}
\end{figure}

For two vertices, following the above considerations, the dominant contribution in $N$, which is again $\Od(N)$, is obtained by taking the two vertices as the two mass effective vertices of Figure \ref{fig:massvertex}. The corresponding connected diagram is shown in Figure \ref{fig:parfun} (d).  Thus, those effective vertices count as $1/N$ each, the external bubbles connected to the vertex also count as $N$, and the internal bubble connecting the two vertices counts an additional $N=\delta_{ab}\delta^{ab}$ coming from $\pif^2(x)\pif^2(y)$. Any other combination is subleading, including the derivative vertices in (\ref{NLSM}). Hence, even if the $\pif^2$  contractions in $f(x)$ are taken maximally in $N$, since $\partial\pif_a \pif^a$ cannot be contracted at the same point  one ends for diagram (d) in Fig. \ref{fig:parfun} with the structure $N\delta^{ab}\delta_{ac}\delta_b^c=N^2$ for one mass vertex and one derivative vertex and $\delta^{ab}\delta_{ac}\delta_{bd}\delta^{cd}=N$ for two derivative vertex, instead of the $N^3$ structure of the two mass vertices. Following the same arguments, other topologies of diagrams considered in the ChPT expansion \cite{Gerber:1988tt} are subdominant for large $N$ at finite mass, like diagram (a) in Figure \ref{fig:otherdiagrams}.

To calculate the dominant $z_2$ contribution from Figure \ref{fig:parfun} (d),  we have to include a combinatoric factor accounting for all the possible ways to choose a $\pif^2$ in each vertex that is to be connected with the other vertex. Thus, at each vertex there will be an additional $k_i+2$ factor, where $k_i$ is the integer labelling the power $(\pif^2)^{k_i+2}$ at vertex $i=1,2$. Hence, we end up with a modified effective vertex function for that diagram, namely:

\begin{eqnarray}
\frac{M^2}{NF^2}\sum_{k=0}^\infty (-1)^k (k+2)\begin{pmatrix} 1/2\cr k+2 \end{pmatrix} x^k=-\frac{1}{2}\frac{M^2}{NF^2}\frac{\tilde g(x)}{x}, \nonumber\\
\tilde g(x)=\frac{1}{4}\frac{d}{dx}\left[x^2 g(x)\right]=-2h'(x)=\frac{1}{\sqrt{1-x}}-1.
\label{modeffvert}
\end{eqnarray}
In addition, we have to multiply  by  2 from the two ways to contract $\pi^2(x)\pi^2(y)$ and by $1/2$ from the Lagrangian expansion, so that the leading $\Od(N)$ term for $z_2$ is:

\begin{eqnarray}
z_2 (M,T)=-\frac{NM^4}{4}\left[\tilde g \left(\frac{G_1(M,T)}{F^2}\right)\right]^2 G_2(M,T) + \Od(N^0)=
-\frac{NM^3T}{32\pi}\left[\tilde g\left(\frac{T^2}{12F^2}\right)\right]^2 + \Od(M^4\log M,N^0)
\label{z2}
\end{eqnarray}
with $G_2(M,T)$ in \eqref{G2} and using the expansions \eqref{g1} and \eqref{g2} around the massless limit.

Now, for three or more vertices $n$, following the previous arguments, the dominant connected diagrams are those with only effective mass vertices and $n+1$ bubbles. Those bubbles can be either closed around one of the vertices or connecting a subset of them, so that  two given vertices are connected at most with two lines.  Examples of these  general $\Od(N)$   foam or super-daisy type diagrams \cite{Dolan:1973qd,Drummond:1997cw} are diagram (e) in Figure \ref{fig:parfun} and diagrams (b) and (c) in Figure \ref{fig:otherdiagrams}. 

However, a crucial point of our present approach is that only a subclass of those foam diagrams is dominant to leading order in the $M^2$ expansion, namely the ring or daisy diagrams of the type depicted in Figure \ref{fig:parfun}.  To understand this, note first that any of those generic diagrams contains a product of the type $G_1^{k_1}G_2^{k_2}\cdots G_n^{k_n}$, multiplied by the corresponding effective vertices (which are functions of $G_1$ only), with $\displaystyle \sum_{i=1}^n k_i=n+1$ the number of loops and $G_k$ given in \eqref{Gk}. Hence, we see that the minimum power of $M$ in a foam diagram is obtained when setting $k_1$ to its maximum allowed value, which is $k_1=n$ and hence $k_n=1$, which corresponds to the loop function $G_n$  connecting those $n$ single bubbles, and the other $k_i=0$. Including also the effective vertices described above, that combination corresponds to the ring diagrams depicted in Figure \ref{fig:parfun}(e), whose $M^2$ counting is then $(M^2)^n G_n=\Od(TM^3)$ times a function of $T^2/12 F^2$, near the chiral limit. Other foam diagrams are subleading, like diagram (b) in Figure \ref{fig:otherdiagrams}, whose leading order mass dependence comes from $(M^2)^3 G_2^2$ and then it becomes $\Od(M^4)$ times a function of $T^2/12 F^2$. Similar arguments can be followed  for other subleading diagrams  like diagram (c) in Figure \ref{fig:otherdiagrams}, whose leading $M$ power is set by the functions $G_{k>1}$. We recall  that the dominance of ring diagrams near the infrared region is a known feature of Thermal Field Theory \cite{Kapusta:2006pm} and allows us precisely to extract the  $\Od(TM^3)$ term needed to calculate the leading order of the scalar susceptibility. 

In addition, and as announced above, to this leading order  all the results are finite, since the $T=0$ corrections are subleading. Therefore, at this level of approximation we will not need to discuss the renormalization details of the calculation in terms of the infinite set of coupling constants arising from higher order Lagrangians \cite{Dobado:1992jg,Dobado:1994fd,Cortes:2015emo}. Within that standard large-$N$ approach, we can consider then that those low-energy constants of higher order are subleading.  We will just keep in mind that they may introduce  subleading corrections to our expressions, which for instance  could  modify numerically the transition temperature, as we discuss below.

The energy density up to $\Od(TM^3)$ in the large-$N$ limit is obtained then by summing all ring diagrams, taking into account that  we have to multiply by: i) $\left[-M^2 \tilde g(G_1/F^2)/(2N)\right]^n$ from the modified effective vertex  \eqref{modeffvert}, counting all the ways to take in each effective vertex the two pion lines shared with the adjacent vertices, and where we have taken into account that there is  a tadpole function $G_1$  attached to every vertex in the ring diagrams,  ii) by  $2^n$ for the two possible ways to take those shared lines in every link, iii) by $(1/n!)$ from the series expansion of the interaction Lagrangian, iv) by $(n-1)!/2$, which are the topologically different ways of sorting  $n$ points in a circle, and v) by $N^{n+1}$ from the loops. Finally we obtain:

\begin{eqnarray}
z(M,T)&=&z_c+\sum_{n=0}^\infty z_n(M,T)=  -N
\frac{\pi^2 T^4}{90}
-NM^2F^2\left\{ 1-\frac{T^2}{24F^2}+h\left(\frac{T^2}{12F^2}\right)
\right\}
 \nonumber\\
 &-& \frac{NM^3T}{8\pi}\left\{\frac{2}{3}-2h' \left(\frac{T^2}{12F^2}\right)+H\left[-\frac{1}{2} \tilde g\left(\frac{T^2}{12F^2}\right)\right]\right\}
  +\Od\left[M^4\log M, N^0\right],
\label{zorderm3}
\end{eqnarray}
 where $\displaystyle H(x)=x^2+2\sum_{n=3}^\infty  \frac{(2n-5)!!}{n!} x^n=-\frac{2}{3}\left(1-3x-\sqrt{1-2x}+2x\sqrt{1-2x}\right)$. The combinatoric factor given here resembles that obtained for a finite-temperature effective Higgs potential in \cite{Arnold:1992rz}.

We remark that the result \eqref{zorderm3} is meaningful only as a expansion around the chiral limit $M\rightarrow 0^+$ so  we can extract the leading order for the quark condensate and the scalar susceptibility in that limit.  Although the mass expansion performed here could yield mass corrections to the free energy, quark condensate and scalar susceptibility, we must be careful at this point.  The $M$ expansion that we have carried out here comes from two different dimensionless parameters, namely $M/T$ and $M/F$. Thus, while we expect  the  behaviour of thermal functions to be dominated by the chiral limit contribution in the high temperature regime $T>>M$,  $M/F$ is a pure $T=0$ parameter coming from the mass vertex in the Lagrangian. Therefore, mass corrections might not be reliable for physical masses.  A related complication introduced by the large-$N$ resummation is that, as showed above, from $z_2$ onwards the effective vertex gives rise actually to $M^2 \tilde g (G_1/F^2)$, with $\tilde g$ given in  \eqref{modeffvert}, which may introduce spurious divergences near $T_c$ when keeping a finite mass $M$.  For these reasons, we will stick here to the strict chiral limit $M\rightarrow 0^+$ since it ensures that those two dimensionless scales can be treated as perturbatively equivalent.  Thus, we will keep only the  $M^0$ term in \eqref{condchiral} for the condensate and the $M^{-1}$ term for the scalar susceptibility. In this work we are interested in the critical behaviour of the large-$N$ expansion and, therefore, it makes sense to restrict to the chiral limit as a first approximation to the problem.  A proper treatment of finite mass effects within the large-$N$ expansion would require in principle to sum all types of foam diagrams like that in Figure \ref{fig:otherdiagrams} (c).

\section{Quark Condensate and Scalar Susceptibility: Results and discussion}
\label{sec:res}

From the analysis of the previous section, we can easily extract the quark condensate in the chiral and large-$N$ limits, from \eqref{quarkconddef} and \eqref{zorderm3}:

\begin{equation}
\frac{\cond(M,T)}{\cond(M,0)}=\sqrt{1-\frac{T^2}{T_c^2}}+\Od(M,1/N).
\label{condchiral}
\end{equation}
where we denote by $T_c^2=12F^2$ the temperature at which the quark condensate vanishes, and $\cond(M,0)=-2NF^2B_0+\Od(M^2\log M)$.  Note that, according to our previous arguments, only the $M^2$ term in \eqref{zorderm3} contributes to the leading order displayed in \eqref{condchiral}. The  result \eqref{condchiral} is certainly expected from the chiral limit analysis in \cite{Bochkarev:1995gi}, where the same temperature dependence is  found  for the scaling of the effective sigma field $\langle \sigma \rangle (T)/\langle \sigma \rangle (0)$. Although the identification $\langle \sigma\rangle \leftrightarrow \cond$ is very natural from the viewpoint of the quark model assignment, note that within our present approach we do not need to introduce any such $\sigma$ field, since we have the exact leading mass dependence of the free energy.  The above result is also numerically compatible with the condensate obtained from the large-$N$ framework in \cite{MeyersOrtmanns:1993dw}, which relies on a saddle point approximation developed via the auxiliary field method and  depends on a cutoff parameter. It agrees also with the result obtained to leading order in $1/N$ within the vector $O(N)$ model in  \cite{Patkos:2002xb,Andersen:2004ae}.

\begin{figure}
\centering
\includegraphics[scale=0.5]{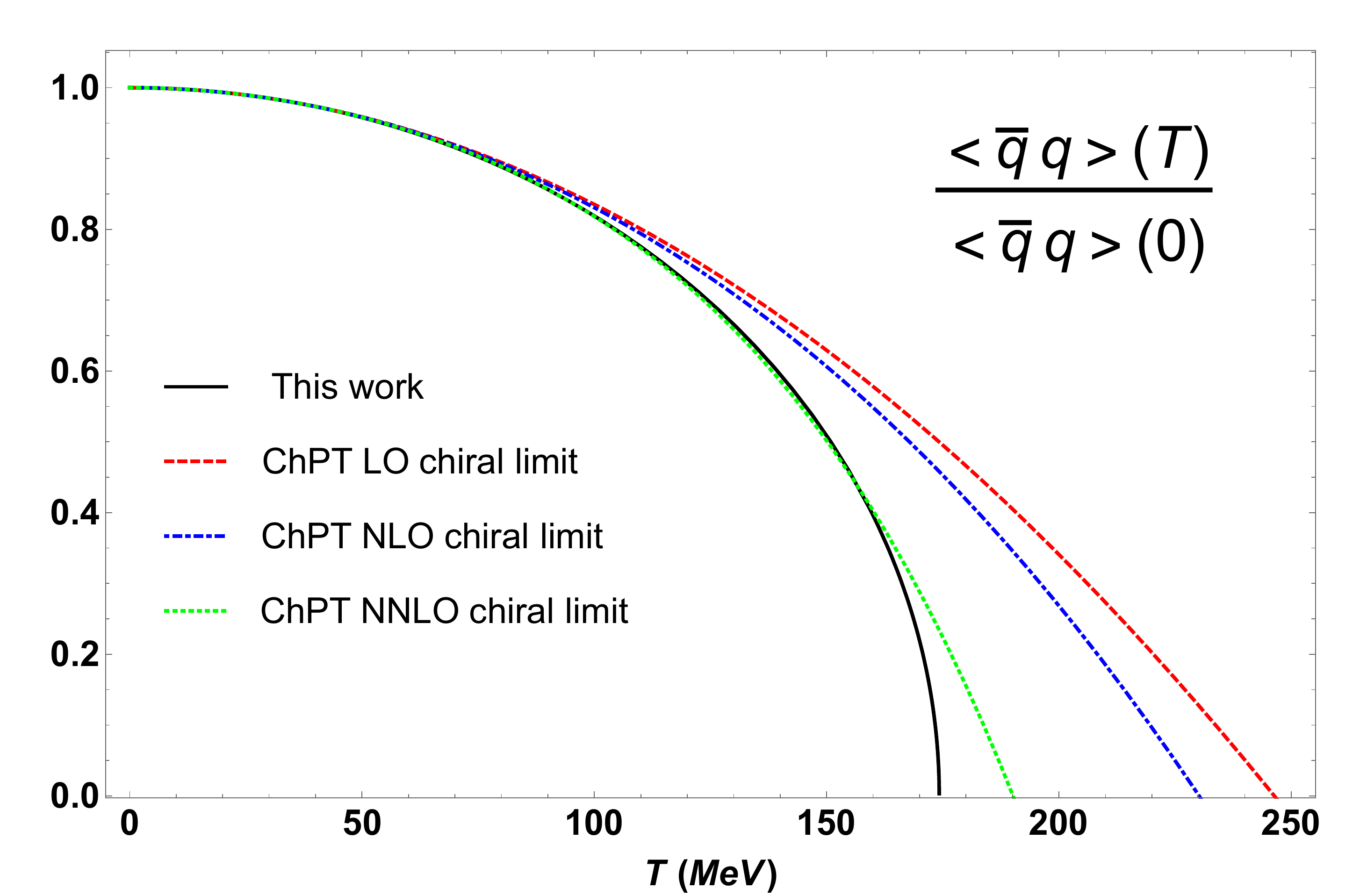}
\caption{Quark condensate in the large-$N$ in the chiral limit compared to the ChPT results for different orders as obtained in \cite{Gerber:1988tt}.}
\label{fig:cond}
\end{figure}

Therefore, the quark condensate at this level of approximation vanishes at $T_c$ and, more importantly, is not defined above $T_c$, unlike the ChPT expression which is just a polynomial whose leading order is recovered just by expanding \eqref{condchiral} in powers of $T^2/T_c^2$, namely $\cond(T)/\cond (0)=1-T^2/(8F^2)+\dots$;  this corresponds to  the leading order ChPT result in the chiral limit for $N=3$ \cite{Gerber:1988tt,Bochkarev:1995gi}. Note also that, consistently with previous studies \cite{Bochkarev:1995gi} $T_c$ is independent of $N$ to leading order. 
 As for the particular value of the transition temperature $T_c$, taking the standard value for the pion decay constant in the chiral limit $\sqrt{N}F\simeq$ 87.1 MeV yields $T_c\simeq $ 174.24 MeV, which is high compared to the restoration temperature in the chiral limit expected from lattice data, which according to our comments in the introduction should be typically around $T_c\sim 120$ MeV. However, the present approach is not meant to provide a $T_c$ close to such lattice values, mostly because is based only on NGB degrees of freedom. Heavier states in the partition function would contribute significantly to reduce the condensate and $T_c$ \cite{Gerber:1988tt,Tawfik:2005qh,GomezNicola:2012uc}. In addition, higher order Lagrangians could modify the $T=0$ value of the condensate and hence the critical temperature. What is more meaningful is to compare the result \eqref{condchiral} with the standard ChPT analysis to different orders in the chiral limit \cite{Gerber:1988tt}, as we  show in Figure \ref{fig:cond}. The value of $T_c$ is  reduced with respect to the NNLO ChPT  while kept within the expected uncertainty given by a $1/N$ expansion near $N=3$. Actually, from the chiral expansion of 
\eqref{condchiral} we get already a factor of $\sqrt{2}$ reduction in $T_c$ with respect to leading order ChPT.

As discussed in the Introduction, the main improvement of the large-$N$ analysis with respect to the ChPT one is that the system undergoes a second-order phase transition,  as corresponds to QCD in the chiral limit, which is clear from the analytic expression \eqref{condchiral} and from Figure \ref{fig:cond}.  The system does not undergo a first-order phase transition since there are not jump discontinuities in the order parameter; the latter behaviour is seen for instance when considering auxiliar field methods to analyze finite-temperature effects in an $O(4)$ nonlinear model \cite{Seel:2011ju}. This second-order critical behaviour will be confirmed by our analysis of the scalar susceptibility below.

Despite the fact that we cannot access from this approach the region $T>T_c$, let us analyze further the analytic behaviour of the condensate \eqref{condchiral} below $T_c$.  At this point it is  important to recall the critical behaviour observed in lattice QCD. In  \cite{Ejiri:2009ac}, the scaling with the quark mass and the temperature of lattice data is  fitted reasonably well to the three-dimensional $O(2)$ and $O(4)$ universality classes (see \cite{Pelissetto:2000ek} for a review). Thus, the quark condensate is expected to scale in the chiral limit  $m_q=0$ as the second-order behaviour $\cond(T)\sim \vert T_c-T\vert^\beta$ for $T\rightarrow T_c^{-}$, where the value of the critical exponent $\beta$ is given in Table \ref{tab:critexp} for the cases considered in \cite{Ejiri:2009ac}. Our value is then $\beta=1/2$ from \eqref{condchiral}, which is not far from  the lattice observations, still within the $1/N$ expected error, although giving a stronger critical behaviour. Note that, as commented above,  we are including only Goldstone bosons and not heavier states and hence we must consider our framework as a qualitative description of the transition, which actually captures its main critical features and certainly improves over  standard low-$T$ expansions such as ChPT. In fact, the critical behaviour is meant to be controlled by Goldstone bosons   below the transition \cite{Pelissetto:2000ek}.

\begin{table}
\begin{tabular}{|c|c|c|c}   \hhline{|===|}
	Model & $\beta$ & $\gamma_\chi^-$   \\ \hline
	3D $O(2)$ &0.35& 0.49 \\ \hline
	3D $O(4)$ &0.38& 0.54  \\ \hline
	This work    &0.5 &  0.75 \\ \hline
	 Saturated thermal pole large $N$ \cite{Cortes:2015emo} &  & 0.975\\ \hline
	Saturated thermal pole IAM \cite{Nicola:2013vma} &  & 1.005 \\
	\hhline{|===|}
	\end{tabular}
\caption{Critical exponents for different universality classes and  the present analysis. We include for reference also the result obtained from saturating the susceptibility with the thermal $f_0(500)$ pole as discussed in \cite{Nicola:2013vma,Cortes:2015emo} and in the main text. In that case, we use the results of the Grayer fit in that paper.}
\label{tab:critexp}
\end{table}

In order to gain more insight into the above mentioned critical features, let us now calculate the leading order contribution of the scalar susceptibility $\chi_S$ from the free energy result \eqref{zorderm3}, keeping only the leading term for $M\rightarrow 0^+$.  After taking into account the mass expansion in \eqref{zorderm3} and the  functions involved, we find the following remarkably simple expression:

\begin{equation}
\chi_S(M,T)=\frac{NTB_0^2}{4\pi M}\left(1-\frac{T^2}{T_c^2}\right)^{-3/4}+\Od(\log M,N^0).
\label{suschiral}
\end{equation}

Expression \eqref{suschiral}  diverges below $T_c$ as a second-order phase transition, as QCD in the chiral limit and confirming our previous analysis of the condensate. Actually, the critical behaviour with $T$ and $M$ for $O(2)$ and $O(4)$ 3D models is given by  $\chi_S(M,T)\sim M^{-1} (T_c-T)^{-\gamma_\chi^-}$ for $T\rightarrow T_c^-$ and $M\rightarrow 0^+$, where the values of the critical exponent $\gamma_\chi^-$ are given in Table \ref{tab:critexp}. Therefore, we reproduce the expected mass behaviour for $\chi_S$ and we obtain a critical exponent slightly above but not far from those models,  which reproduce fairly well the critical behaviour of lattice QCD \cite{Ejiri:2009ac}. As explained above, this is more than reasonable for a description based only on the lightest NGB which again lies within the expected numerical uncertainty while capturing the main features of the chiral transition.  Note also that, consistently, from \eqref{suschiral} we recover the leading order ChPT expression for the thermal part of the susceptibility in the chiral limit \cite{Smilga:1995qf,GomezNicola:2012uc} as $\chi_S=\frac{NTB_0^2}{4\pi M}+\dots$.  The $T=0$ vanishes at this order and is included in the $\log M$ neglected corrections in \eqref{suschiral}.

\begin{figure}
\centering
\includegraphics[scale=0.6]{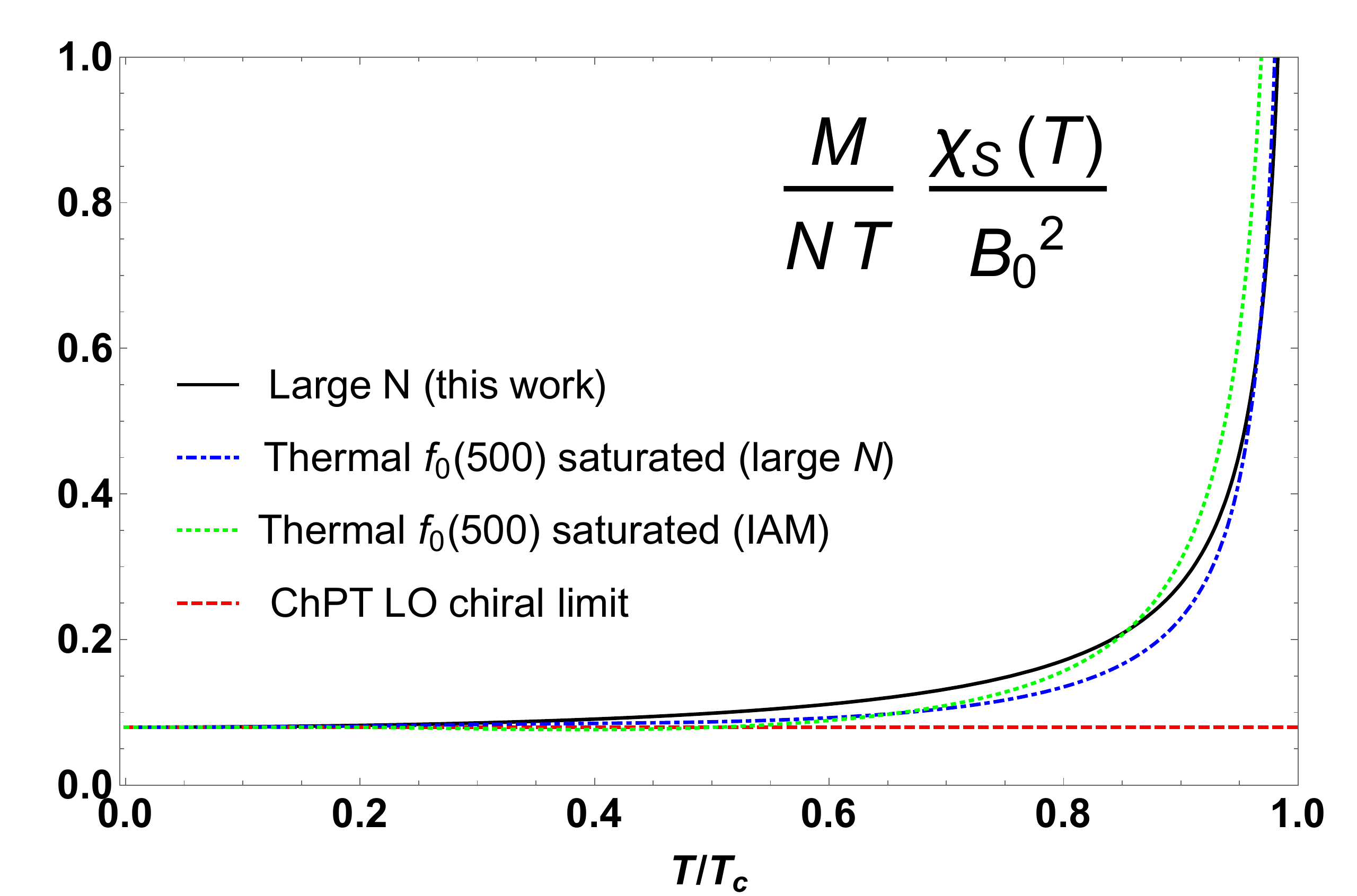}
\caption{Scalar susceptibility in the present approach compared to the thermal $f_0(500)$ saturated one defined through \eqref{chisat}, considered in \cite{Cortes:2015emo} at large $N$, where the results with Grayer fit in that paper have been taken. We also include the saturated susceptibility with the IAM \cite{Nicola:2013vma} in the chiral limit and the ChPT LO chiral limit result for comparison. Temperatures are rescaled to the critical temperature for each case, namely $T_c=174.24$ MeV for the present approach, $T_c=92.33$ MeV for the saturated large-$N$ one, $T_c=118.2$ MeV for the saturated IAM and $T_c=246.42$ MeV for the ChPT chiral limit one.}
\label{fig:suscep}
\end{figure}

Finally, we also compare our results with the analysis performed in \cite{Nicola:2013vma} and \cite{Cortes:2015emo}, from which one can define a scalar susceptibility saturated by the thermal $f_0(500)$ state, with a mass corresponding to that scalar resonance defined as $M_S^2(T)=M_p^2(T)-\Gamma_p^2(T)/4$, where $s_p=(M_p-i\Gamma_p/2)^2$ is the position of the pole in the second Riemann sheet of the scattering partial wave with isospin and angular momentum $I=J=0$ calculated also at finite temperature within the large-$N$ limit in \cite{Cortes:2015emo} and with the IAM in \cite{Nicola:2013vma} around the chiral limit. In order that such saturated susceptibility complies properly with the expected low-$T$ behaviour given by the LO ChPT in the chiral limit, we define it as 

\begin{equation}\chi_S^{sat}(T)=\frac{NTB_0^2}{4\pi M}\frac{M_S^2(0)}{M_S^2(T)}\label{chisat}\end{equation}
neglecting the $T=0$ logarithmic contribution near $M\rightarrow 0^{+}$, and we plot $M \chi_S/(B_0^2 NT)$ for different approaches, which is then $N$-independent. The results are showed in Figure \ref{fig:suscep}, where we also include the ChPT chiral limit one for comparison, which to leading order is just a constant with the normalization chosen. Higher order ChPT corrections yield smoothly growing functions of $T$. In the saturated large-$N$ case, we have used the parameters of the so called Grayer fit in \cite{Cortes:2015emo}.

First of all, we observe that the values of the critical temperature differ considerably between the free-energy analysis, like the one we present here or ChPT, and those based on thermal $f_0(500)$ saturation (either IAM or large-$N$). As we have commented above, the numerical value of $T_c$ obtained from the present approach is presumably affected by higher order terms and does not account for any physical state other than the NGB ones. In contrast, the saturation approach incorporates successfully the thermal $f_0(500)$, relying on a good description of the  physical $T=0$ pole consistent with scattering data and the quoted PDG values for that state. In fact, the critical temperature of that approach is  closer to the expected values from lattice analysis, which highlights the importance of that thermal state. Nevertheless, the most important result here has to do with the critical behaviour, so that in order to provide a   clearer comparison of the two approaches in that sense,  we have represented the susceptibility in terms of $T/T_c$, where $T_c$ is different for each method.

As it can be seen from  Figure \ref{fig:suscep}, although the critical exponent for the saturated $\chi^{sat}(T)$ are larger (the numerical results are given also in Table \ref{tab:critexp} \footnote{In the saturated cases, the critical exponents $\gamma^-$ in Table \ref{tab:critexp} differ from those quoted in  \cite{Cortes:2015emo} by small numerical corrections which come from taking more points close to the transition temperature in the present work.}), they depart later from the low-$T$ ChPT value, so in the end the two approaches remain very close near the critical region. This is an important check of consistency between those two different ways to determine $\chi_S$, concerning its critical behaviour.

\section{Conclusions}

In this work we have analyzed the quark condensate and the scalar susceptibility  of a gas of $N$ Goldstone Bosons to leading order in $N$ in the chiral limit. To obtain the leading  behavior of the susceptibility requires to compute up to $M^3$ corrections in the free energy (which come from an infinite set of closed ring diagrams). To this order, the results are directly finite and then, it is no necessary to add higher order Lagrangian counterterms within the usual large-$N$ framework. This diagrammatic treatment in terms of thermal effective vertices and dominant diagrams  would allow us to extend this analysis beyond the chiral limit or for higher orders in the $1/N$ expansion. 

Our results show a critical behaviour in reasonable agreement with the universality classes expected for lattice simulations of the chiral transition, both for the quark condensate and for the scalar susceptibility, within the numerical uncertainties expected for a large-$N$ approach.  This is particularly realized in the critical exponents for those quantities. The quark condensate improves over NNLO ChPT in the sense that it reduces the value of its vanishing point and, more importantly, it behaves as the order parameter of a second-order transition, the critical $T_c$ being $N$-independent to leading order. The differences in the critical temperature with respect to ChPT remain within the expected $1/N$ uncertainty. Likewise, the scalar susceptibility that we have calculated here diverges at the same $T_c$ as the condensate vanishing, as it should.  In fact, another motivation of the present work was to test the consistency of a recent approach based on saturating the scalar susceptibility with the thermal $f_0(500)$ resonance pole. Our present analysis shows a reasonable agreement between the two methods regarding the critical behaviour in terms of $T/T_c$. However, the value of $T_c$ is much lower in the saturated approaches that in those based on the partition function, like ChPT or the one we present here. This is most likely due to having incorporated properly the physical $f_0(500)$ state and its thermal dependence, which somehow mimics the influence of higher order contributions in the perturbative chiral approach.

To obtain higher order corrections in the mass expansion would require to sum additional infinite sets of diagrams, the so-called foam diagrams. This would be especially interesting for the case of the scalar susceptibility, since in principle it should lead  to a change  from a divergent second-order transition to a smooth peak characteristic of a crossover, as observed in lattice simulations. That analysis is beyond the scope of this work and will be analyzed elsewhere. Nevertheless, we believe that the present study can be useful as a first approach to this problem, setting up the diagrammatic framework for future studies while extending previous analysis in a nontrivial way and capturing the main features of chiral restoration in QCD.

\section*{Acknowledgments}
Work partially supported by contracts FPA2014-53375-C2-2-P and FIS2014-57026-REDT (Spanish Hadron Excelence Network) from the Spanish ``Ministerio de Econom\'ia y Competitividad". We also acknowledge the support
of the HadronPhysics3 project from the European Union Seventh Framework Programme (EU-FP7). Santiago Cort\'es thanks Prof. Jos\'e Rolando Rold\'an and the High Energy Physics group of Universidad de los Andes and Colombian `Departamento Administrativo de Ciencia, Tecnologia e Innovaci\'on" (COLCIENCIAS) for  financial support.

\end{document}